\documentclass[prd,draft,preprint,showpacs,groupedaddress]{revtex4-1}
\usepackage{amsmath}
\usepackage{amsfonts}
\usepackage{amssymb}
\usepackage{geometry}
\usepackage{natbib}
\usepackage[english]{babel}
\usepackage{indentfirst}
\usepackage{mathrsfs}

\begin{document}

  \setlength{\parindent}{2em}
  \title{Apparent horizon and gravitational thermodynamics of the Universe in Eddington-Born-Infeld theory}
  \author{Jia-Cheng Ding} \author{Qi-Qi Fan} \author{Cong Li} \author{Ping Li}
  \author{Jian-Bo Deng} \email[Jian-Bo Deng: ]{dengjb@lzu.edu.cn}
  \affiliation{Institute of Theoretical Physics, LanZhou University, Lanzhou 730000, P. R. China}
  \date{\today}

  \begin{abstract}

  The thermodynamics of the Universe in Eddington-Born-Infeld (EBI) theory is restudied by utilizing the holographic-style gravitational equations that dominate the dynamics of the cosmical apparent horizon $\Upsilon_{A}$ and the evolution of Universe. We started from applying the Bigravity method to rewrite the EBI action of the Palatini approach into the Bigravity-type action with an extra metric $q_{\mu\nu}$. With the help of the holographic-style dynamical equations, we discussed the properties of the cosmical apparent horizon $\Upsilon_{A}$ including timelike, spacelike and null characters, which depend on the value of the parameter of state $w_{m}$ in EBI Universe. We also obtained the unified first law for the gravitational thermodynamics and the total energy differential for the open system enveloped by $\Upsilon_{A}$ in EBI Universe. Finally, we used the Gibbs equation in the positive-heat-out sign convention to derive the generalized second laws of the nondecreasing entropy $S_{tot}^{(A)}$ enclosed by $\Upsilon_{A}$ in EBI universe.

  \end{abstract}



  \maketitle

  \section{Introduction}

  The thermodynamics of the Universe is quite an interesting problem that has attracted a lot of researchers. Recently, many studies have covered both the first and second laws of thermodynamics for the Friedmann-Robertson-Walker (FRW) Universe with a generic spatial curvature. The inspired work  is the first law of thermodynamics for the Universe by Cai and Kim~\cite{1}, which is a part of the effort to seek the connections between thermodynamics and gravity~\cite{2} after discovering the black-hole thermodynamics~\cite{3,4}. In Ref.~\cite{1,5}, Akbar and Cai reversed the formulation by rewriting the Friedmann equations into the heat balance equation and the unified first law of thermodynamics at the cosmical apparent horizon, for general relativity (GR), Gauss-Bonnet and Lovelock gravities. The results of Ref.~\cite{5} was soon generalized to other theories of gravity, such as the scalar-tensor gravity~\cite{6}, $f(R)$ gravity~\cite{7}, braneworld scenarios~\cite{8,9,10}, generic $f(R, \phi, \nabla_{\mu}\phi\nabla^{\mu}\phi)$ gravity~\cite{11}, and Horava-Lifshitz gravity~\cite{12,13}, to construct the effective total energy differentials by the corresponding modified Friedmann equations.
\par
  The Eddington-Born-Infeld action (EBI) was proposed in Ref.~\cite{15}, which could mimic the presence of dark energy and dark matter~\cite{15,16,18} in the expansion of the Universe and modify the Newton-Poisson equation that leads to flat rotation curves for galaxies. Someone proposed that the EBI action was a candidate for nonparticulate dark matter and dark energy. Generally, the EBI action is a Palatini-type action where the metric $g_{\mu\nu}$ is not associate with the connection $C_{\mu\nu}^{\lambda}$. However by defining an extra metric $q_{\mu\nu}$ that satisfies the condition about $g\mu\nu$ and $C_{\mu\nu}^{\lambda}$ in Ref.~\cite{15,17}, the EBI action can be rewritten as the Bigravity-type action, which is our starting point.
\par
  Inspired by the gravitational thermodynamics in these gravitational theories~\cite{5,6,7,8,9,10,11,12,13,14} and characteristics of the EBI action, we focused on generalizing the results in Ref.~\cite{6,14} to the EBI gravity. We derived the holographic-style dynamical equations and discussed the properties of the cosmical apparent horizon $\Upsilon_{A}$ in EBI Universe, which rely on the contents of the cosmical apparent horizon including the matter and the dark energy provided by the cosmological constant $\Lambda$ and the spacetime self-coupling. Furthermore, we applied the Misner-Sharp energy, the Cai-Kim temperature $\hat{T}_{A}$ and the Hawking-Bekenstein entropy $S_{A}$ to obtain the unified first law for the gravitational thermodynamics and the total energy differential for the open system enveloped by $\Upsilon_{A}$ in EBI Universe. Finally, we derived the generalized second laws of the nondecreasing entropy $S_{tot}^{(A)}$ enclosed by $\Upsilon_{A}$ in the EBI universe.
\par
  This paper is organized as follows. In Sec.\uppercase\expandafter{\romannumeral2}, we reviewed the cosmical apparent horizon and derived the holographic-style dynamical equations in the EBI theory. Then we discussed the properties of the cosmical apparent horizon. In Sec.\uppercase\expandafter{\romannumeral3}, the unified first laws of gravitational thermodynamics and the Clausius equation on $\Upsilon_{A}$ for an isochoric process in EBI Universe were discussed. We also derived the total energy differential enclosed by $\Upsilon_{A}$ in EBI Universe. In Sec.\uppercase\expandafter{\romannumeral4}, the generalized second law of gravitational thermodynamics in the EBI Universe was derived. Conclusions and discussion are given in Sec.\uppercase\expandafter{\romannumeral5}.

  \section{Dynamics of the cosmical apparent horizon in Eddington\,-\,Born\,-\,Infeld gravity}

  \subsection{Apparent horizon}

  Physically, apparent horizons constitute the observable boundary which is the largest boundary of Universe in an instant. Mathematically, apparent horizons are many hypersurfaces where the outward expansion rate $\theta_{(\ell)}$ or the inward expansion rate $\theta_{(n)}$ is equal to zero. In general, the first kind of apparent horizons, corresponding to that $\theta_{(\ell)}=0$ and $\theta_{(n)}\neq0$, usually locate near the black holes and the another kind of apparent horizon, corresponding to that $\theta_{(n)}=0$ and $\theta_{(\ell)}\neq0$, appear in the vicinity of the expanding boundary of Universe, called the cosmical apparent horizons. In this paper, we only discussed the cosmical apparent horizon via dynamic equations of Universe and thermodynamic methods.
\par
  In order to calculate the apparent horizon of the cosmology, one used the FRW metric to describe the spatially homogeneous and isotropic Universe~\cite{1,19}
  \begin{equation}
  \begin{aligned}
    \label{eq:The line element of FRW metric}
    ds^2 &=-dt^2+\frac{a(t)^2}{1-kr^2}dr^2+a(t)^2r^2(d\theta^2+\sin^2\theta d\varphi^2)
  \end{aligned}
  \end{equation}
  where $a(t)$ is the scale factor of the evolution of Universe and the index $k$ denotes the normalized spatial curvature, with $k$ = \{+1, 0, -1\} corresponding to closed, flat, and open Universes, respectively. Using the spherical symmetry, the metric can be rewritten as
  \begin{equation}
  \begin{aligned}
    \label{eq:The spherical symmetric line element of the FRW metric}
    ds^2=h_{\alpha\beta}dx^{\alpha}dx^{\beta}+\Upsilon^2(d\theta^2+\sin^2\theta d\varphi^2)
  \end{aligned}
  \end{equation}
  where $h_{\alpha\beta}:=diag[-1, \frac{a(t)^2}{1-kr^2}]$ represents the transverse 2-metric spanned by ($x^{0}=t, x^{1}=r$)and $\Upsilon:=a(t)r$ stands for the astronomical circumference/areal radius. Based on the FRW metric, one can structure the following null tetrad adapted to the spherical symmetry and the null radial flow:
   \begin{equation}
   \begin{aligned}
    \label{eq:The null tetrad }
   \ell^{\mu}&=\frac{1}{\sqrt{2}}(1, \frac{\sqrt{1-kr^2}}{a}, 0, 0)\\
   n^{\mu}&=\frac{1}{\sqrt{2}} (-1, \frac{\sqrt{1-kr^2}}{a}, 0, 0)\\
   m^{\mu}&=\frac{1}{\sqrt{2}\Upsilon}(0, 0, 1, \frac{i}{\sin\theta})\\
   \bar{m}^{\mu}&=\frac{1}{\sqrt{2}\Upsilon}(0, 0, 1, -\frac{i}{\sin\theta}),
   \end{aligned}
   \end{equation}
  corresponding to the metric signature $(-, +, +, +)$. By calculating the Newman-Penrose spin coeffcients $\rho_{NP}:=-m^{\mu}\bar{m}^\nu\nabla_{\nu}\ell_{\mu}$ and $\mu_{NP}:=\bar{m}^{\mu}m^{\nu}\nabla_{\nu}n_{\mu}$ , the outward expansion rate $\theta_{(\ell)}=-(\rho_{NP}+\bar{\rho}_{NP})$ and the inward expansion rate $\theta_{(n)}=-(\mu_{NP}+\bar{\mu}_{NP})$ are, respectively, given by
  \begin{equation}
   \begin{aligned}
    \label{eq:The outward and inward expansion rates }
    \theta_{(\ell)}&=\sqrt{2}[H+\Upsilon^{-1}\sqrt{1-\frac{k\Upsilon^{2}}{a^{2}}}]\\
    \theta_{(n)}&=\sqrt{2}[-H+\Upsilon^{-1}\sqrt{1-\frac{k\Upsilon^{2}}{a^2}}],
   \end{aligned}
   \end{equation}
  where $H:=\frac{\dot{a}}{a}$ is the Hubble parameter of cosmic spatial expansion. The overdot denotes the derivative with respect to the comoving time $t$.
\par
  For the expanding Universe $(H>0)$, the cosmical apparent horizon is given by
  \begin{equation}
  \label{eq:The cosmical apparent horizon equation}
   \Upsilon_{A}=\frac{1}{\sqrt{H^2+\frac{k}{a^2}}},
  \end{equation}
  derived from $\theta_{(n)}=0$ and $\theta_{(\ell)}>0$ corresponding to the unique marginally inner trapped horizon where $\partial_{\mu}\Upsilon$ becomes a null vector with $g^{\mu\nu}\partial_{\mu}\Upsilon\partial_{\nu}\Upsilon=0$~\cite{20}. Then one derives the temporal derivative of Eq.\eqref{eq:The cosmical apparent horizon equation}
  \begin{equation}
  \label{eq:The kinematic equation of the cosmical apparent horizon}
   \dot{\Upsilon}_{A}=-H\Upsilon_{A}^3(\dot{H}-\frac{k}{a^2})
  \end{equation}
  that is a kinematic equation of the cosmical apparent horizon.

  \subsection{The holographic-style dynamical equations in Eddington\,-\,Born\,-\,Infeld Universe}

  The action of Eddington-Born-Infeld theory is given by ~\cite{15,16,17,18}
  \begin{equation}
  \begin{aligned}
  \label{eq:The action of Eddington-Born-Infeld theory}
  S_{EBI}(g_{\mu\nu},\, C^{\mu}_{\nu\rho},\,\mathcal{L}_{m})=&\frac{1}{16\pi G}\int d^{4}x\{\sqrt{-g}(R-2\Lambda)+\frac{2}{\alpha\ell^{2}}\sqrt{-det(g_{\mu\nu}-\ell^{2}K_{\mu\nu}(C))}\}\\
  &+\int d^4x\sqrt{-g}\mathcal{L}_{m},
  \end{aligned}
  \end{equation}
  where $R$ is the Ricci scalar for the metric $g_{\mu\nu}$ and $g$ represents the determinant of $g_{\mu\nu}$. $K_{\mu\nu}$ is two-order Riemann curvature tensor dependent with the connection $C^{\mu}_{\nu\rho}$, provided by the Palatini approach, which has no concern with $g_{\mu\nu}$. $\Lambda$ is the cosmological constant and $\alpha$ is an arbitrary constant. $G$ is the gravitational constant and $\mathcal{L}_{m}$ is the Lagrangian density of matter.
\par
   We applying the Bigravity method~\cite{17} to replace the connection $C^{\mu}_{\nu\rho}$ by the extra metric $q_{\mu\nu}$ the EBI theory, the action \eqref{eq:The action of Eddington-Born-Infeld theory} can be rewritten into the Bigravity-type action
  \begin{equation}
  \begin{aligned}
  \label{eq:The action of Bigravity from the EBI theory}
  S_{EBI}=\frac{1}{16\pi G}&\int d^{4}x\{\sqrt{-g}(R-2\Lambda)+\sqrt{-q}(K-2\lambda)-\frac{1}{\ell^{2}}\sqrt{-q}(q^{\alpha\beta}g_{\alpha\beta})\}\\
  &+\int d^4x\sqrt{-g}\mathcal{L}_{m},
  \end{aligned}
  \end{equation}
  where
  \begin{equation}
  \begin{aligned}
  \label{eq:The relation between q and K}
  q_{\mu\nu}=-\frac{1}{\alpha}(g_{\mu\nu}-\ell^{2}K_{\mu\nu}).
  \end{aligned}
  \end{equation}
  $K_{\mu\nu}$ is the Ricci tensor for the extra metric $q_{\mu\nu}$ and $K$ is the Ricci scalar for the extra metric $q_{\mu\nu}$. $\lambda$ is a constant ($\lambda\equiv\frac{\alpha}{\ell^{2}}$) corresponding to $q_{\mu\nu} $ and $q$ is the determinant of $q_{\mu\nu}$. Here, both $g_{\mu\nu}$ and $q_{\mu\nu}$ are innate metrics of spacetime and they are mutually independent. Hence, $\frac{1}{\ell^{2}}\sqrt{-q}(q^{\alpha\beta}g_{\alpha\beta})$ can be regard as the term from self-coupling of spacetime.
\par
  Varying the Bigravity-type action \eqref{eq:The action of Bigravity from the EBI theory} with respect to the metric $g_{\mu\nu}$ yields the field equations~\cite{15}
  \begin{equation}
  \begin{aligned}
  \label{eq:The field equations in the EBI theory}
  R_{\mu\nu}-\frac{1}{2}R g_{\mu\nu}=8\pi G T_{\mu\nu}^{(m)}-\Lambda g_{\mu\nu}-\frac{1}{\ell^{2}}\frac{\sqrt{-q}}{\sqrt{-g}}g_{\mu\alpha}q^{\alpha\beta}g_{\beta\nu},
  \end{aligned}
  \end{equation}
  where $-\frac{1}{\ell^{2}}\frac{\sqrt{-q}}{\sqrt{-g}}g_{\mu\alpha}q^{\alpha\beta}g_{\beta\nu}$ is the energy-momentum tensor of the spacetime self-coupling.
\par
  The matter content of the Universe is construed as the perfect fluid whose the energy-momentum tensor is
  \begin{equation}
  \begin{aligned}
  \label{eq:The energy-momentum tensor of the perfect fluid}
   T^{\mu\;(m)}_{\;\;\nu}&=diag[-\rho_{m}, p_{m}, p_{m}, p_{m}]\\
   &with \  \frac{p_{m}}{\rho_{m}}=:w_{m},
  \end{aligned}
  \end{equation}
  where $w_{m}$ refers to the Equation-of-State(EoS) parameter of the perfect fluid. In order to study the cosmological property of the EBI Universe, we made $g_{\mu\nu}$ become the FRW metric and assumed the extra metric $q_{\mu\nu}$~\cite{15,35} as
  \begin{equation}
  \begin{aligned}
  \label{eq:The line element of the extra metric}
  ds_{q}^{2}=-Udt^{2}+\frac{a(t)^{2}V}{1-kr^{2}}dr^{2}+a(t)^{2}Vr^{2}d\theta^{2}+a(t)^{2}Vr^{2}\sin^{2}\theta d\varphi^{2}.
  \end{aligned}
  \end{equation}
  $U$ and $V$ are two undetermined positive functions independent with $t$, which are corresponding to the intensity of the spacetime self-coupling.
\par
  Depending on the the field equation and two metrics, one can get the first Friedmann equation
  \begin{equation}
  \begin{aligned}
  \label{eq:The first Friedmann equation of the EBI gravity}
  \frac{\dot{a}^{2}}{a^{2}}+\frac{k}{a^{2}}-\frac{\Lambda}{3}=\frac{8\pi G\rho_{m}}{3}+\frac{1}{3\ell^{2}}\sqrt{\frac{V}{U}}V
  \end{aligned}
  \end{equation}
  and the second Friedmann equation
  \begin{equation}
  \begin{aligned}
  \label{eq:The second Friedmann equation of the EBI gravity}
  \frac{\dot{a}^{2}}{a^{2}}+\frac{k}{a^{2}}+2\frac{\ddot{a}}{a}-\Lambda=-8\pi G p_{m}+\frac{1}{\ell^{2}}\sqrt{\frac{V}{U}}U.
  \end{aligned}
  \end{equation}
  Eq.\eqref{eq:The second Friedmann equation of the EBI gravity} can be rewritten into
  \begin{equation}
  \begin{aligned}
  \label{eq:The second Friedmann equation in another form of the EBI gravity}
  2\dot{H}+3H^{2}+\frac{k}{a^{2}}-\Lambda=-8\pi G p_{m}+\frac{1}{\ell^{2}}\sqrt{\frac{V}{U}}U,
  \end{aligned}
  \end{equation}
  which is equivalent to
  \begin{equation}
  \begin{aligned}
  \label{eq:The third holographic-style dynamical equation}
  \Upsilon_{A}^{-3}(\dot{\Upsilon}_{A}-\frac{3}{2}H\Upsilon_{A})=[4\pi G p_{m}-\frac{\Lambda}{2}-\frac{1}{2\ell^{2}}\sqrt{\frac{V}{U}}U]H.
  \end{aligned}
  \end{equation}
\par
  With the help of Eq.\eqref{eq:The first Friedmann equation of the EBI gravity} and Eq.\eqref{eq:The second Friedmann equation in another form of the EBI gravity}, we obtain
  \begin{equation}
  \begin{aligned}
  \label{eq:014}
  \dot{H}-\frac{k}{a^{2}}=-4\pi G (\rho_{m}+p_{m})-\frac{1}{2\ell^{2}}\sqrt{\frac{V}{U}}(V-U).
  \end{aligned}
  \end{equation}
  Based on Eq.\eqref{eq:The cosmical apparent horizon equation} and Eq.\eqref{eq:The first Friedmann equation of the EBI gravity}, we get
  \begin{equation}
  \begin{aligned}
  \label{eq:The first holographic-style dynamical equation}
  \Upsilon_{A}^{-2}=\frac{8\pi G}{3}\rho_{m}+\frac{\Lambda}{3}+\frac{1}{3\ell^{2}}\sqrt{\frac{V}{U}}V
  \end{aligned}
  \end{equation}
  and substituting Eq.\eqref{eq:014} into Eq.\eqref{eq:The kinematic equation of the cosmical apparent horizon}, we get
  \begin{equation}
  \begin{aligned}
  \label{eq:The second holographic-style dynamical equation}
  \dot{\Upsilon}_{A}=H\Upsilon_{A}^{3}[4\pi G(\rho_{m}+p_{m})+\frac{1}{2\ell^{2}}\sqrt{\frac{V}{U}}(V-U)].
  \end{aligned}
  \end{equation}
\par
  Eq.\eqref{eq:The third holographic-style dynamical equation}, Eq.\eqref{eq:The first holographic-style dynamical equation} and Eq.\eqref{eq:The second holographic-style dynamical equation} constitute the holographic-style dynamical equations about the cosmical apparent horizon~\cite{14}, which means the evolution of Universe enveloped by $\Upsilon_{A}$ can be described by the characters of the cosmical apparent horizon. If one takes $U=V=1$ and $\Lambda=0$, the holographic-style dynamical equations will return to the condition of the Einstein theory.
\par
  Further more, from Eq.\eqref{eq:The first Friedmann equation of the EBI gravity} and \eqref{eq:The second Friedmann equation of the EBI gravity}, we can obtain the acceleration equation of the EBI Universe
   \begin{equation}
   \begin{aligned}
   \label{eq:The accelerated equation of the scalar factor 'a' in the EBI Universe}
   \frac{\ddot{a}}{a}&=-\frac{4\pi G}{3}(\rho_{m}+3p_{m})+\frac{\Lambda}{3}-\frac{1}{2\ell^{2}}\sqrt{\frac{V}{U}}(\frac{1}{3}V-U)\\
   &=-4\pi G\rho_{m}[w_{m}+\frac{1}{3}-\frac{\Lambda}{12\pi G \rho_{m}}+\frac{1}{8\pi G\rho_{m}\ell^{2}}\sqrt{\frac{V}{U}}(\frac{1}{3}V-U)].
   \end{aligned}
   \end{equation}

  \subsection{The characters of the cosmical apparent horizon}

  In general, the cosmical apparent horizon is not null surface, which is different from the event and particle horizon. The equation of the cosmical apparent horizon in comoving coordinates is~\cite{27}
  \begin{equation}
  \begin{aligned}
  \label{eq:The constraint equation of the apparent horizon}
  \mathscr{F}(t, r)=a(t)r-\frac{1}{\sqrt{H^{2}+\frac{k}{a^{2}}}}=0.
  \end{aligned}
  \end{equation}
  Its normal has components
  \begin{equation}
  \begin{aligned}
  \label{eq:The components of the normal vector of AH}
  N_{\mu}&=\nabla_{\mu}\mathscr{F}|_{AH}=\{[\dot{a}r+\frac{H(\dot{H}-\frac{k}{a^{2}})}{(H^{2}+\frac{k}{a^{2}})^{\frac{3}{2}}}]\delta_{\mu0}+a\delta_{\mu1}\}|_{AH}\\
  &=H\Upsilon_{A}[1+(\dot{H}-\frac{k}{a^{2}})\Upsilon_{A}^{2}]\delta_{\mu0}+a\delta_{\mu1}\\
  &=H\Upsilon_{A}^{3}\frac{\ddot{a}}{a}\delta_{\mu0}+a\delta_{\mu1}.
  \end{aligned}
  \end{equation}
  The norm squared of the normal vector is
  \begin{equation}
  \begin{aligned}
  \label{eq:The norm squared of the normal vector of AH}
  N_{a}N^{a}&=1-k r_{A}^{2}-H^{2}\Upsilon_{A}^{6}(\frac{\ddot{a}}{a})^{2}\\
  &=H^{2}\Upsilon_{A}^{2}[1-\Upsilon_{A}^{4}(\frac{\ddot{a}}{a})^{2}]\\
  &=H^{2}\Upsilon^{6}_{A}(\Upsilon^{-2}_{A}-\frac{\ddot{a}}{a})(\Upsilon^{-2}_{A}+\frac{\ddot{a}}{a}),
  \end{aligned}
  \end{equation}
  where $r_{A}=\frac{\Upsilon_{A}}{a}$.
  Substituting Eq.\eqref{eq:The accelerated equation of the scalar factor 'a' in the EBI Universe} and  Eq.\eqref{eq:The first holographic-style dynamical equation},  we get
  \begin{equation}
  \begin{aligned}
  \label{eq:}
  N_{a}N^{a}&=\mathscr{H}(w_{m})=-H^{2}\Upsilon^{6}_{A}(4\pi G\rho_{m})^{2}[w_{m}-\frac{1}{3}-\frac{\Lambda}{6\pi G\rho_{m}}\\
  &-\frac{1}{8\pi G\rho_{m}\ell^{2}}\sqrt{\frac{V}{U}}(\frac{1}{3}V+U)][w_{m}+1+\frac{1}{8\pi G\rho_{m}\ell^{2}}\sqrt{\frac{V}{U}}(V-U)],
  \end{aligned}
  \end{equation}
  where we consider that $N_{a}N^{a}$ is only the quadratic function $\mathscr{H}(w_{m})$ representing the inner product of the normal vector of the cosmical apparent horizon. The quadratic function $\mathscr{H}(w_{m})$ has two zero points, $w_{m}=\frac{1}{3}+\frac{\Lambda}{6\pi G\rho_{m}}+\frac{1}{8\pi G\ell^{2}\rho_{m}}\sqrt{\frac{V}{U}}(\frac{1}{3}V+U)$ and $w_{m}=-[1+\frac{1}{8\pi G\ell^{2}\rho_{m}}\sqrt{\frac{V}{U}}(V-U)]$.
\par
  Considering the properties of the quadratic function $\mathscr{H}(w_{m})$, we get three results as follows. ( Here we consider the condition that $\Lambda>-\frac{1}{\ell^{2}}\sqrt{\frac{V}{U}}V-8\pi G \rho_{m}$. )\\
  \indent A. when $w_{m}=\frac{1}{3}+\frac{\Lambda}{6\pi G\rho_{m}}+\frac{1}{8\pi G\ell^{2}\rho_{m}}\sqrt{\frac{V}{U}}(\frac{1}{3}V+U)$ or $w_{m}=-[1+\frac{1}{8\pi G\ell^{2}\rho_{m}}\sqrt{\frac{V}{U}}(V-U)]$, $N_{a}N^{a}=0$ that shows the normal vector $N^{a}$ is a null vector and the apparent horizon $\Upsilon_{A}$ is a null surface. It coincides with the cosmological event horizon $\Upsilon_{E}=a\int_{t}^{\infty}a^{-1}d\hat{t}$, which by definition is a future-pointed null causal boundary~\cite{22,27}. And it shares the signature of isolated black-hole horizons~\cite{23}.\\
  \indent B. when $-[1+\frac{1}{8\pi G\ell^{2}\rho_{m}}\sqrt{\frac{V}{U}}(V-U)]<w_{m}<[\frac{1}{3}+\frac{\Lambda}{6\pi G\rho_{m}}+\frac{1}{8\pi G\ell^{2}\rho_{m}}\sqrt{\frac{V}{U}}(\frac{1}{3}V+U)]$, $N_{a}N^{a}>0$ that shows $N^{a}$ is a spacelike vector and $\Upsilon_{A}$ is the timelike surface. $\Upsilon_{A}$ has the signature $(-, +, +)$ that shares the signature of a quasilocal timelike membrane in black-hole physics~\cite{24,25}.\\
  \indent C. when $[\frac{1}{3}+\frac{\Lambda}{6\pi G\rho_{m}}+\frac{1}{8\pi G\ell^{2}\rho_{m}}\sqrt{\frac{V}{U}}(\frac{1}{3}V+U)]<w_{m}$ or $w_{m}<-[1+\frac{1}{8\pi G\ell^{2}\rho_{m}}\sqrt{\frac{V}{U}}(V-U)]$, $N_{a}N^{a}<0$ that shows $N^{a}$ is a timelike vector and $\Upsilon_{A}$ is the spacelike surface. Its signature is $(+, +, +)$ that is same with the signature of the dynamical black hole horizons~\cite{26}.\\
\par
    As we know, the Universe is accelerated expanding that means the matter outside the cosmical apparent horizon may enter into the cosmical apparent horizon with the evolution of the Universe. Hence we considered that the timelike cosmical apparent horizon is reasonable and the range of the EoS parameter $-[1+\frac{1}{8\pi G\ell^{2}\rho_{m}}\sqrt{\frac{V}{U}}(V-U)]<w_{m}<[\frac{1}{3}+\frac{\Lambda}{6\pi G\rho_{m}}+\frac{1}{8\pi G\ell^{2}\rho_{m}}\sqrt{\frac{V}{U}}(\frac{1}{3}V+U)]$ is significative, which is similar with the range of the EoS parameter ($-1<w<\frac{1}{3}$) in Einstein Universe~\cite{14}.

   \section{Thermodynamics of the holographic-style dynamical equations in the Eddington\,-\,Born\,-\,Infeld Universe }

   Based on the holographic-style dynamical equations\eqref{eq:The first holographic-style dynamical equation}, \eqref{eq:The second holographic-style dynamical equation} and \eqref{eq:The third holographic-style dynamical equation} in Sec.\uppercase\expandafter{\romannumeral2}, we continue to investigate the thermodynamics about the cosmical apparent horizon. We define the total energy within a sphere of radius $\Upsilon$, surface area $A=4\pi \Upsilon^{2}$, and volume $\hat{V}=\frac{4}{3}\pi \Upsilon^{3}$: $E_{tot}=\rho_{tot}\hat{V}$. (We take $\hat{V}$ to represent the volume in order to distinguish the function $V$.)

   \subsection{Unified first law of thermodynamics}

   Applying the Misner-Sharp mass/energy $E_{MS}:=\frac{\Upsilon}{2G}(1-h^{\alpha\beta}\partial_{\alpha}\Upsilon\partial_{\beta}\Upsilon)$~\cite{28,29} to be the total energy and substituting $h_{\alpha\beta}=diag[-1, \frac{a^{2}}{1-kr^{2}}]$, one obtain
   \begin{equation}
   \begin{aligned}
   \label{eq:026}
   dE=-\frac{\dot{\Upsilon}_{A}}{G}\frac{\Upsilon^{3}}{\Upsilon_{A}^{3}}dt+\frac{3}{2G}\frac{\Upsilon^{2}}{\Upsilon_{A}^{2}}d\Upsilon
   \end{aligned}
   \end{equation}
   and
   \begin{equation}
   \begin{aligned}
   \label{eq:027}
   dE=-\frac{1}{G}\frac{\Upsilon^{3}}{\Upsilon_{A}^{3}}(\dot{\Upsilon}_{A}-\frac{3}{2}H\Upsilon_{A})dt + \frac{3}{2G}\frac{\Upsilon^{2}}{\Upsilon_{A}^{2}}adr.
   \end{aligned}
   \end{equation}
   For the EBI Universe, substituting Eq.\eqref{eq:The third holographic-style dynamical equation} and Eq.\eqref{eq:The first holographic-style dynamical equation} into Eq.\eqref{eq:026} yields
   \begin{equation}
   \begin{aligned}
   \label{eq:The total energy differential in (t,Y) coordinates}
   dE=&A[\rho_{m}+\frac{1}{8\pi G\ell^{2}}\sqrt{\frac{V}{U}}V+\frac{\Lambda}{8\pi G}]d\Upsilon-A[(\rho_{m}+p_{m})\\
   &+\frac{1}{8\pi G\ell^{2}}\sqrt{\frac{V}{U}}(V-U)]\cdot H\Upsilon dt,
   \end{aligned}
   \end{equation}
   which is called the total energy differential in the $(t, \Upsilon)$ coordinates, where $A=4\pi \Upsilon^{2}$.
   Similarly, we can obtain
   \begin{equation}
   \begin{aligned}
   \label{eq:The total energy differential in (t,r) coordinates}
   dE=A[\rho_{m}+\frac{1}{8\pi G\ell^{2}}\sqrt{\frac{V}{U}}V+\frac{\Lambda}{8\pi G}]adr-A[p_{m}-\frac{1}{8\pi G \ell^{2}}\sqrt{\frac{V}{U}}U-\frac{\Lambda}{8\pi G}]\cdot H\Upsilon dt,
   \end{aligned}
   \end{equation}
   which is the total energy differential in the $(t, r)$ coordinates. From the above, we can know that the energy density $\rho_{sc}$ provided by spacetime self-coupling is $\frac{1}{8\pi G\ell^{2}}\sqrt{\frac{V}{U}}V$ and the intensity of pressure $p_{sc}$ provided by spacetime self-coupling is $-\frac{1}{8\pi G \ell^{2}}\sqrt{\frac{V}{U}}U$, which indicates the EoS parameter $w_{ac}$, provided by the dark energy from spacetime self-coupling, is negative.
\par
   The unified first law of (equilibrium) thermodynamics is given by
   \begin{equation}
   \begin{aligned}
   \label{eq:The unifed first law in Einstein theory (Hayward)}
   dE=A \boldsymbol{\Psi}+W d\hat{V},
   \end{aligned}
   \end{equation}
   proposed by Hayward ~\cite{30}.
   $W$ is the work density, given by
   \begin{equation}
   \begin{aligned}
   \label{eq:The definition of the work density}
   W:=-\frac{1}{2}T_{(m)}^{\alpha\beta}h_{\alpha\beta}
   \end{aligned}
   \end{equation}
   where $h_{\alpha\beta}=diag[-1, \frac{a(t)^{2}}{1-kr^{2}}]$.
   $\boldsymbol{\Psi}$ is the energy supply covector, $\boldsymbol{\Psi}=\boldsymbol{\Psi}_{\alpha}dx^{\alpha}$, where
   \begin{equation}
   \begin{aligned}
   \label{eq:The definition of the component of the energy supply covector}
   \boldsymbol{\Psi}_{\alpha}:=T_{\alpha(m)}^{\beta}\partial_{\beta}\Upsilon+W\partial_{\alpha}\Upsilon.
   \end{aligned}
   \end{equation}
   Here, $W$ and $\boldsymbol{\Psi}_{\alpha}$ is invariant.
   Moreover, the definitions of $W$ and $\boldsymbol{\Psi}_{\alpha}$ are valid for all spherically symmetric spacetimes and FRW spacetime.
\par
   In the EBI theory, the field equation can be rewritten into
   \begin{equation}
   \begin{aligned}
   \label{eq:The field equations with the total energy-momentum tensor}
   R_{\mu\nu}-\frac{1}{2}R g_{\mu\nu}=8\pi G T_{\mu\nu}^{(tot)},
   \end{aligned}
   \end{equation}
   where we define a total energy-momentum tensor
   \begin{equation}
   \begin{aligned}
   \label{eq:The definition of the total energy-momentum tensor}
   T_{\mu\nu}^{(tot)}=T_{\mu\nu}^{(m)}-\frac{\Lambda}{8\pi G}g_{\mu\nu}-\frac{1}{8\pi G \ell^{2}}\sqrt{\frac{V}{U}}UV\cdot g_{\mu\alpha}q^{\alpha\beta}g_{\beta\nu}.
   \end{aligned}
   \end{equation}
   From the above equation, we can consider that the total energy-momentum tensor includes the part of the dark energy corresponding to the terms with $\Lambda$ and $(U,V)$. Then we can generalize the Hayward's unified first law of (equilibrium) thermodynamics by taking $T_{\mu\nu}^{(tot)}$ to replace $T_{\mu\nu}^{(m)}$. Imitating the definitions of $W$ and $\boldsymbol{\Psi}_{\alpha}$~\cite{30}, we define
   \begin{equation}
   \begin{aligned}
   \label{eq:The unifed first law in EBI theory}
   dE=A \boldsymbol{\tilde{\Psi}}+\tilde{W} d\hat{V}
   \end{aligned}
   \end{equation}
   and
   \begin{equation}
   \begin{aligned}
   \label{eq:The expression of the energy supply covector}
   \boldsymbol{\tilde{\Psi}}=\boldsymbol{\tilde{\Psi}}_{\alpha}dx^{\alpha},
   \end{aligned}
   \end{equation}
   where
   \begin{equation}
   \begin{aligned}
   \label{eq:The definition of the work density in EBI theory}
   \tilde{W}:=-\frac{1}{2}T_{(tot)}^{\alpha\beta}h_{\alpha\beta}
   \end{aligned}
   \end{equation}
   and
   \begin{equation}
   \begin{aligned}
   \label{eq:The definition of the component of the energy supply covector in EBI theory}
   \tilde{\boldsymbol{\Psi}}_{\alpha}:=T_{\alpha \,(tot)}^{\beta}\partial_{\beta}\Upsilon+\tilde{W}\partial_{\alpha}\Upsilon.
   \end{aligned}
   \end{equation}
\par
   We consider that the FRW metric $g_{\mu\nu}$ is physically subsistent that can be observed, which is used to raise or descend the index here, and another metric $q_{\mu\nu}$ is an extra metric, provided by the primordial mechanism of Universe, which can be considered as a correction of the Einstein theory.  So, we use $g_{\mu\nu}$ to raise $T^{(tot)}_{\alpha\beta}$:
   \begin{equation}
   \begin{aligned}
   \label{eq:T tot up B down a}
   T_{\alpha (tot)}^{\beta}&=g^{\mu\beta}T_{\mu\alpha}^{(tot)}\\
   &=g^{\mu\beta}T_{\mu\alpha}^{(m)}-\frac{\Lambda}{8\pi G}\delta_{\alpha}^{\beta}-\frac{1}{8\pi G\ell^{2}}\sqrt{\frac{V}{U}}UV q^{\beta\mu}g_{\mu\alpha}
   \end{aligned}
   \end{equation}
   and
   \begin{equation}
   \begin{aligned}
   \label{eq:T tot up a up B}
   T^{\alpha\beta}_{(tot)}&=g^{\mu\alpha}g^{\nu\beta}T_{\mu\nu}^{(tot)}\\
   &=g^{\mu\alpha}g^{\nu\beta}T_{\mu\nu}^{(m)}-\frac{\Lambda}{8\pi G}g^{\alpha\beta}-\frac{1}{8\pi G\ell^{2}}\sqrt{\frac{V}{U}}UV q^{\alpha\beta}.
   \end{aligned}
   \end{equation}
   Substituting $h_{\alpha\beta}=diag[-1, \frac{a(t)^{2}}{1-kr^{2}}]$, we obtain
   \begin{equation}
   \begin{aligned}
   \label{eq:The calculation of the work density in EBI theory}
   \tilde{W}&=-\frac{1}{2}[T^{00}_{(tot)}h_{00}+T^{11}_{(tot)}h_{11}]\\
   &=\frac{1}{2}(\rho_{m}-p_{m})+\frac{\Lambda}{8\pi G}+\frac{1}{8\pi G\ell^{2}}\sqrt{\frac{V}{U}}(\frac{U+V}{2});
   \end{aligned}
   \end{equation}
   \begin{equation}
   \begin{aligned}
   \label{eq:The calculation of the 't' component of the energy supply covector in EBI theory in the (t, r) coordinates}
   \indent\tilde{\boldsymbol{\Psi}}_{t}=-\frac{1}{2}[(\rho_{m}+p_{m})+\frac{1}{8\pi G\ell^{2}}\sqrt{\frac{V}{U}}(V-U)]H\Upsilon;
   \end{aligned}
   \end{equation}
   \begin{equation}
   \begin{aligned}
   \label{eq:The calculation of the 'r' component of the energy supply covector in EBI theory in the (t, r) coordinates}
   \tilde{\boldsymbol{\Psi}}_{r}=\frac{1}{2}[(\rho_{m}+p_{m})+\frac{1}{8\pi G\ell^{2}}\sqrt{\frac{V}{U}}(V-U)]a.
   \end{aligned}
   \end{equation}
   Substituting $\tilde{\boldsymbol{\Psi}}_{t}$, $\tilde{\boldsymbol{\Psi}}_{r}$ and $\tilde{W}$ into Eq.\eqref{eq:The unifed first law in EBI theory}, we get
   \begin{equation}
   \begin{aligned}
   \label{eq:The result of the unifed first laws in EBI theory in the (t, r) coordinates}
   dE&=A\boldsymbol{\tilde{\Psi}}+\tilde{W}d\hat{V}\\
   &=A[\boldsymbol{\tilde{\Psi}}_{t}dt+\boldsymbol{\tilde{\Psi}}_{r}dr+\tilde{W}d\Upsilon]\\
   &=A[-p_{m}+\frac{\Lambda}{8\pi G}+\frac{1}{8\pi G\ell^{2}}\sqrt{\frac{V}{U}}U]H\Upsilon dt+A[\rho_{m}+\frac{\Lambda}{8\pi G}+\frac{1}{8\pi G\ell^{2}}\sqrt{\frac{V}{U}}V]adr,
   \end{aligned}
   \end{equation}
   which is the expression of $dE$ in $(t, r)$ coordinates.
\par
   Naturally, because of the invariance of $\tilde{W}$ and $\tilde{\boldsymbol{\Psi}}$, we can rewrite these in the $(t, \Upsilon)$ coordinates, given by
   \begin{equation}
   \begin{aligned}
   \label{eq:The result of the unifed first laws in EBI theory in the (t, Y) coordinates}
   dE&=A\boldsymbol{\tilde{\Psi}}+\tilde{W}d\hat{V}\\
   &=A[\boldsymbol{\tilde{\Psi}}'_{t}dt+\boldsymbol{\tilde{\Psi}}'_{\Upsilon}d\Upsilon+\tilde{W}d\Upsilon]\\
   &=-A[(\rho_{m}+p_{m})+\frac{1}{8\pi G\ell^{2}}\sqrt{\frac{V}{U}}(V-U)]H\Upsilon dt\\
   &+A[\rho_{m}+\frac{\Lambda}{8\pi G}+\frac{1}{8\pi G\ell^{2}}\sqrt{\frac{V}{U}}V]d\Upsilon,
   \end{aligned}
   \end{equation}
   where
   \begin{equation}
   \begin{aligned}
   \label{eq:The calculation of the 't' component of the energy supply covector in EBI theory in the (t, Y) coordinates}
   \indent\tilde{\boldsymbol{\Psi}}'_{t}=-[(\rho_{m}+p_{m})+\frac{1}{8\pi G\ell^{2}}\sqrt{\frac{V}{U}}(V-U)]H\Upsilon
   \end{aligned}
   \end{equation}
   and
   \begin{equation}
   \begin{aligned}
   \label{eq:The calculation of the 'Y' component of the energy supply covector in EBI theory in the (t, Y) coordinates}
   \tilde{\boldsymbol{\Psi}}'_{\Upsilon}=\frac{1}{2}[(\rho_{m}+p_{m})+\frac{1}{8\pi G\ell^{2}}\sqrt{\frac{V}{U}}(V-U)].
   \end{aligned}
   \end{equation}
\par
   It illustrates our hypothesis of replacing $T_{\mu\nu}^{(m)}$ by $T_{\mu\nu}^{(tot)}$ is reasonable that Eq.\eqref{eq:The result of the unifed first laws in EBI theory in the (t, r) coordinates} and Eq.\eqref{eq:The result of the unifed first laws in EBI theory in the (t, Y) coordinates} are respectively identical to Eq.\eqref{eq:The total energy differential in (t,r) coordinates} and Eq.\eqref{eq:The total energy differential in (t,Y) coordinates}. The unified first laws for gravitational thermodynamics of Universe are totally different from the first laws in the black-hole thermodynamics~\cite{14}. Eq.\eqref{eq:The result of the unifed first laws in EBI theory in the (t, r) coordinates} and Eq.\eqref{eq:The result of the unifed first laws in EBI theory in the (t, Y) coordinates} are called the unified first laws for the EBI Universe's gravitational thermodynamics.

    \subsection{Clausius equation on the cosmical apparent horizon for an isochoric process}

   Having obtained the unified first law $dE=A\tilde{\boldsymbol{\Psi}}+\tilde{W} d\hat{V}$ in EBI Universe, we are interested in the region enclosed by the cosmical apparent horizon $\Upsilon_{A}$.
   \par
   Eq.\eqref{eq:The second holographic-style dynamical equation} leads to
   \begin{equation}
   \begin{aligned}
   \label{eq:048}
   \frac{\dot{\Upsilon}_{A}}{G}dt=A_{A}[(\rho_{m}+p_{m})+\frac{1}{8\pi G\ell^{2}}\sqrt{\frac{V}{U}}(V-U)]H\Upsilon_{A}\cdot dt,
   \end{aligned}
   \end{equation}
   where $A_{A}=4\pi\Upsilon_{A}^{2}$ and the left-hand side can be manipulated into ~\cite{14}
   \begin{equation}
   \begin{aligned}
   \label{eq:049}
   \frac{\dot{\Upsilon}_{A}}{G}dt=\frac{1}{2\pi \Upsilon_{A}}\cdot(\frac{2\pi \Upsilon_{A}\dot{\Upsilon}_{A}}{G}\cdot dt)=\frac{1}{2\pi \Upsilon_{A}}d(\frac{\pi\Upsilon_{A}^{2}}{G}).
   \end{aligned}
   \end{equation}
   One applies the geometrically defined Hawking-Bekenstein entropy ~\cite{4,31} (in the units $\hbar=c=k[Boltzmann\;constant]=1$)
   \begin{equation}
   \begin{aligned}
   \label{eq:The Hawking-Bekenstein entropy}
   S_{A}=\frac{\pi \Upsilon_{A}^{2}}{G}=\frac{A_{A}}{4G}
   \end{aligned}
   \end{equation}
   that is the entropy of the cosmical apparent horizon rather than the inner entropy enclosed by $\Upsilon_{A}$ and the Cai-Kim temperature ~\cite{1,21}
   \begin{equation}
   \begin{aligned}
   \label{eq:The Cai-Kim temperature}
   \hat{T}_{A}\equiv\frac{1}{2\pi\Upsilon_{A}},
   \end{aligned}
   \end{equation}
   at the cosmical apparent horizon to simplify Eq.\eqref{eq:049}, given by $\frac{\dot{\Upsilon}_{A}}{G}dt=\hat{T}_{A}d S_{A}$ (here we take ``$\hat{T}$'' on behalf of temperature not only the Cai-Kim temperature). With the help of Eq.\eqref{eq:The calculation of the 't' component of the energy supply covector in EBI theory in the (t, Y) coordinates}, we get $-[(\rho_{m}+p_{m})+\frac{1}{8\pi G\ell^{2}}\sqrt{\frac{V}{U}}(V-U)]H\Upsilon_{A}=\tilde{\boldsymbol{\Psi}}'_{t}|_{(\Upsilon=\Upsilon_{A})}\equiv\tilde{\boldsymbol{\Psi}}'_{tA}$. From Eq.\eqref{eq:The result of the unifed first laws in EBI theory in the (t, Y) coordinates} and Eq.\eqref{eq:048}, we obtain
   \begin{equation}
   \begin{aligned}
   \label{eq:The Clausius equation for equilibrium and reversible thermodynamic processes}
   \delta Q_{A}=\hat{T}_{A}dS_{A}=-A_{A} \tilde{\boldsymbol{\Psi}}'_{tA} =-dE_{A}|_{d\Upsilon=0},
   \end{aligned}
   \end{equation}
   where $dE_{A}$ represents the condition in $\Upsilon=\Upsilon_{A}$ of Eq.\eqref{eq:The result of the unifed first laws in EBI theory in the (t, Y) coordinates}. Eq.\eqref{eq:The Clausius equation for equilibrium and reversible thermodynamic processes} is actually the Clausius equation for equilibrium and reversible thermodynamic processes as same as the situation in Einstein gravity~\cite{14}. In Einstein Universe, $dE_{A}|_{d\Upsilon=0}$ only contain the part of matter $(\rho_{m}, p_{m})$ without the dark energy. In brief, we considering the EBI theory that is a modified theory of gravity, the Clausius equation on the apparent horizon is generalized to include dark energy with respect to cosmological constant$\Lambda$ and the term of the self-coupling of spacetime $(U, V)$, which may explain the problems about the cosmic expansion.
   \par
   Finally, for the open system enveloped by $\Upsilon_{A}$, we combine the unified first law Eq.\eqref{eq:The result of the unifed first laws in EBI theory in the (t, Y) coordinates} and the Clausius equation \eqref{eq:The Clausius equation for equilibrium and reversible thermodynamic processes} into the total energy differential
   \begin{equation}
   \begin{aligned}
   \label{eq:The total energy differential for open system enveloped by Y-A}
   dE_{A}=-\hat{T}_{A}dS_{A}+[\rho_{m}+\frac{\Lambda}{8\pi G}+\frac{1}{8\pi G\ell^{2}}\sqrt{\frac{V}{U}}V]d\hat{V}_{A},
   \end{aligned}
   \end{equation}
   where the ``-'' sign before ``$\hat{T}_{A}dS_{A}$'' shows the positive-heat-out sign convention that means heat emitted by the open system takes positive values ( $\delta Q_{A}=\delta Q_{A}^{(out)}>0$ ) rather than the traditional positive-heat-in thermodynamic sign convention~\cite{14}.

   Comparing Eq.\eqref{eq:The total energy differential for open system enveloped by Y-A} with the total energy differential of the Einstein gravity, there are two extra terms $\frac{\Lambda}{8\pi G}d\hat{V}_{A}$ and $\frac{1}{8\pi G\ell^{2}}\sqrt{\frac{V}{U}}V d\hat{V}_{A}$ corresponding to dark energy, which originate from the cosmological constant and the spacetime self-coupling respectively. It means that there are the matter's transition and the dark-energy's fluxion on both sides of the cosmical apparent horizon during the expansion of Universe.

   \section{Generalized second laws of thermodynamics in Eddington\,-\,Born\,-\,Infeld Universe }

   With the help of the first holographic-style dynamical equation\eqref{eq:The first holographic-style dynamical equation}, the total energy ($E_{MS}=\frac{\Upsilon^{3}}{2G \Upsilon_{A}^{2}}$) can be rewritten into
   \begin{equation}
   \begin{aligned}
   \label{eq:126}
   E_{MS}=\rho_{tot}(\frac{4}{3}\pi\Upsilon^{3})=\rho_{tot}\hat{V},
   \end{aligned}
   \end{equation}
   where $\rho_{tot}=\rho_{m}+\frac{\Lambda}{8\pi G}+\frac{1}{8\pi G\ell^{2}}\sqrt{\frac{V}{U}}V$.
   Then we can rewrite the Friedmann equations into
   \begin{equation}
   \begin{aligned}
   \label{eq:0126}
   \frac{\dot{a}}{a}+\frac{k}{a}=\frac{8\pi G\rho_{tot}}{3}
   \end{aligned}
   \end{equation}
   and
   \begin{equation}
   \begin{aligned}
   \label{eq:0127}
   \frac{\dot{a}}{a}+\frac{k}{a}+2\frac{\ddot{a}}{a}=-8\pi G p_{tot},
   \end{aligned}
   \end{equation}
   where we define that $p_{tot}=p_{m}-\frac{\Lambda}{8\pi G}-\frac{1}{8\pi G \ell^{2}}\sqrt{\frac{V}{U}}U$. Based on two above equations, one can obtain the continuity equation in the EBI theory
   \begin{equation}
   \begin{aligned}
   \label{eq:The continuity equation in EBI theory}
   \dot{\rho}_{tot}+3\frac{\dot{a}}{a}(\rho_{tot}+p_{tot})=0.
   \end{aligned}
   \end{equation}

\par
   In many papers~\cite{32,33,34}, the entropy $S_{m}$ of the cosmic energy-matter content with temperature $\hat{T}_{m}$ is always determined by the traditional Gibbs equation $dE=\hat{T}_{m}dS_{m}-p_{m}d\hat{V}$. In order to generalize to the EBI theory, we keep the same form and redefine it into the positive-heat-out sign convention for consistency with the horizon entropy $S_{A}$~\cite{14}, given by
   \begin{equation}
   \begin{aligned}
   \label{eq:the Gibbs equation in the "positive-heat-out" sign convention in the EBI theory}
   dE_{tot}=-\hat{T}_{tot}dS_{tot}-p_{tot}d\hat{V}.
   \end{aligned}
   \end{equation}
   From $dE_{tot}=\rho_{tot}d\hat{V}+\hat{V}d\rho_{tot}$, one can obtian
   \begin{equation}
   \begin{aligned}
   \label{eq:061}
   \hat{T}_{tot}dS_{tot}=-\hat{V}d\rho_{tot}-(\rho_{tot}+p_{tot})d\hat{V}.
   \end{aligned}
   \end{equation}
   With the help of the continuity equation\eqref{eq:The continuity equation in EBI theory}, we obtain
   \begin{equation}
   \begin{aligned}
   \label{eq:062}
   d\rho_{tot}=-3H(\rho_{tot}+p_{tot})dt.
   \end{aligned}
   \end{equation}
   When $\Upsilon=\Upsilon_{A}$, one get
   \begin{equation}
   \begin{aligned}
   \label{eq:063}
   \hat{T}_{tot}dS_{tot}^{(A)}=A_{A}(\rho_{tot}+p_{tot})\cdot(H\Upsilon_{A}-\dot{\Upsilon}_{A})dt,
   \end{aligned}
   \end{equation}
   where $A_{A}=\frac{3}{2G \rho_{tot}}$. Based on $\rho_{tot}+p_{tot}=(\rho_{m}+p_{m})+\sqrt{\frac{V}{U}}\frac{(V-U)}{8\pi G\ell^{2}}$, Eq.\eqref{eq:The second holographic-style dynamical equation} can be rewritten into
   \begin{equation}
   \begin{aligned}
   \label{eq:064}
   \dot{\Upsilon}_{A}&=4\pi GH \Upsilon_{A}^{3}(\rho_{tot}+p_{tot})\\
   &=\frac{3}{2}H\Upsilon_{A}\frac{1}{\rho_{tot}}(\rho_{tot}+p_{tot}).
   \end{aligned}
   \end{equation}
   Substituting Eq.\eqref{eq:064} into Eq.\eqref{eq:063} yields
   \begin{equation}
   \begin{aligned}
   \label{eq:The evolution of total entropy enclosed by Y with ( rho and p )}
   \dot{S}_{tot}^{(A)}=-\frac{9}{4G}\cdot\frac{H\Upsilon_{A}}{\hat{T}_{tot}}\cdot\frac{1}{\rho_{tot}^{2}}(\rho_{tot}+p_{tot})(\frac{1}{3}\rho_{tot}+p_{tot})
   \end{aligned}
   \end{equation}
   that is the evolution of the total inner entropy enclosed by $\Upsilon_{A}$.
\par
   We have assumed that $\rho_{tot}\equiv\rho_{m}+\frac{1}{8\pi G\ell^{2}}\sqrt{\frac{V}{U}}V+\frac{\Lambda}{8\pi G}$ and $p_{tot}\equiv p_{m}-\frac{1}{8\pi G\ell^{2}}\sqrt{\frac{V}{U}}U-\frac{\Lambda}{8\pi G}$. In order to discuss $\dot{S}_{tot}^{(A)}$ more concretely, we assume that $\rho_{aux}\equiv\rho_{m}+\frac{1}{8\pi G\ell^{2}}\sqrt{\frac{V}{U}}V$ and $p_{aux}\equiv p_{m}-\frac{1}{8\pi G\ell^{2}}\sqrt{\frac{V}{U}}U$, which are two auxiliary parameters. Following $p_{m}=w_{m}\rho_{m}$, we set
   \begin{equation}
   \begin{aligned}
   \label{eq:0-67}
   p_{tot}=\varepsilon_{tot}\rho_{tot}
   \end{aligned}
   \end{equation}
   and
   \begin{equation}
   \begin{aligned}
   \label{eq:0-68}
   p_{aux}=\sigma_{aux}\rho_{aux},
   \end{aligned}
   \end{equation}
   where $\varepsilon_{tot}$ and $\sigma_{aux}$ are the parameters of state like $w_{m}$. And Eq.\eqref{eq:The evolution of total entropy enclosed by Y with ( rho and p )} can be simplified as
   \begin{equation}
   \begin{aligned}
   \label{eq:The evolution of total entropy enclosed by Y with (varepsilon)}
   \dot{S}_{tot}^{(A)}=-\frac{9}{4G}\cdot\frac{H\Upsilon_{A}}{\hat{T}_{tot}}\cdot(\varepsilon_{tot}+1)(\varepsilon_{tot}+\frac{1}{3}).
   \end{aligned}
   \end{equation}
   \par
   Based on $p_{m}=w_{m}\rho_{m}$ and Eq.\eqref{eq:0-68}, we get
   \begin{equation}
   \begin{aligned}
   \label{eq:0-70}
   \sigma_{aux}=w_{m}+\frac{\sqrt{\frac{V}{U}}(w_{m}V+U)}{8\pi G\ell^{2}\rho_{m}+\sqrt{\frac{V}{U}}V}
   \end{aligned}
   \end{equation}
   and, obviously, there are
   \begin{equation}
   \begin{aligned}
   \label{eq:0-71}
   \varepsilon_{tot}=\sigma_{aux}-(\sigma_{aux}+1)\frac{\Lambda}{8\pi G \rho_{aux}+\Lambda}.
   \end{aligned}
   \end{equation}
   Based on the above equations, we obtain
   \begin{equation}
   \begin{aligned}
   \label{eq:0-72}
   \varepsilon_{tot}=\frac{8\pi G\ell^{2}\rho_{m}}{8\pi G\ell^{2}\rho_{m}+\sqrt{\frac{V}{U}}V+\Lambda\ell^{2}}\cdot w_{m}-\frac{\sqrt{\frac{V}{U}}U+\Lambda\ell^{2}}{8\pi G\ell^{2}\rho_{m}+\sqrt{\frac{V}{U}}V+\Lambda\ell^{2}}.
   \end{aligned}
   \end{equation}
   As a result, $\dot{S}_{tot}^{(A)}$ can be rewritten into
   \begin{equation}
   \begin{aligned}
   \label{eq:The evolution of total entropy enclosed by Y with (w-m)}
   \dot{S}_{tot}^{(A)}&=-\frac{9}{4G}\cdot\frac{H\Upsilon_{A}}{\hat{T}_{tot}}\cdot(\frac{8\pi G\ell^{2}\rho_{m}}{8\pi G\ell^{2}\rho_{m}+\sqrt{\frac{V}{U}}V+\Lambda\ell^{2}})^{2}\cdot[w_{m}+1+\frac{\sqrt{\frac{V}{U}}(V-U)}{8\pi G \ell^{2}\rho_{m}}]\\
   &\cdot[w_{m}+\frac{1}{3}+\frac{\sqrt{\frac{V}{U}}(\frac{1}{3}V-U)}{8\pi G \ell^{2}\rho_{m}}-\frac{\Lambda}{12\pi G \rho_{m}}].
   \end{aligned}
   \end{equation}
   Physically, we consider that temperature is positive ( $\hat{T}_{tot}>0$ ) and the Universe is expanding ( $H>0$ ).
\par
   If $\Lambda$ satisfies the precondition ($\Lambda>-\frac{1}{\ell^{2}}\sqrt{\frac{V}{U}}V-8\pi G \rho_{m}$) in Sec.\uppercase\expandafter{\romannumeral2}, we can exactly obtain that $[1+\frac{\sqrt{\frac{V}{U}}(V-U)}{8\pi G \ell^{2}\rho_{m}}]>[\frac{1}{3}+\frac{\sqrt{\frac{V}{U}}(\frac{1}{3}V-U)}{8\pi G \ell^{2}\rho_{m}}-\frac{\Lambda}{12\pi G \rho_{m}}]$. And Eq.\eqref{eq:The evolution of total entropy enclosed by Y with (w-m)} illustrate that\\
   \indent A. when $-[1+\frac{\sqrt{\frac{V}{U}}(V-U)}{8\pi G \ell^{2}\rho_{m}}]<w_{m}<-[\frac{1}{3}+\frac{\sqrt{\frac{V}{U}}(\frac{1}{3}V-U)}{8\pi G \ell^{2}\rho_{m}}-\frac{\Lambda}{12\pi G \rho_{m}}]$, $\dot{S}_{tot}^{(A)}>0$;\\
   \indent B. when $-[1+\frac{\sqrt{\frac{V}{U}}(V-U)}{8\pi G \ell^{2}\rho_{m}}]>w_{m}$ or $-[\frac{1}{3}+\frac{\sqrt{\frac{V}{U}}(\frac{1}{3}V-U)}{8\pi G \ell^{2}\rho_{m}}-\frac{\Lambda}{12\pi G \rho_{m}}]<w_{m}$, $\dot{S}_{tot}^{(A)}<0$;\\
   \indent C. when $w_{m}=-[1+\frac{\sqrt{\frac{V}{U}}(V-U)}{8\pi G \ell^{2}\rho_{m}}]$ or $-[\frac{1}{3}+\frac{\sqrt{\frac{V}{U}}(\frac{1}{3}V-U)}{8\pi G \ell^{2}\rho_{m}}-\frac{\Lambda}{12\pi G \rho_{m}}]$, $\dot{S}_{tot}^{(A)}=0$.
\par
   On the other hand, from the acceleration equation of the EBI Universe \eqref{eq:The accelerated equation of the scalar factor 'a' in the EBI Universe}, we know that the Universe is accelerated expanding when $w_{m}<-[\frac{1}{3}-\frac{\Lambda}{12\pi G \rho_{m}}+\frac{\sqrt{\frac{V}{U}}(\frac{1}{3}V-U)}{8\pi G \ell^{2}\rho_{m}}]$. If $V<3U-8\pi G\ell^{2}\rho_{m}\sqrt{\frac{U}{V}}$, $w_{m}$ has the possibility to be a positive number to produce a accelerated expanding Universe, which is different from the result of Einstein Universe~\cite{14}.
\par
   In a word, the physical total entropy $S_{tot}^{(A)}$ inside the cosmical apparent horizon satisfies $\dot{S}_{tot}^{(A)}>0$ for the stage of accelerated expansion ($\ddot{a}>0$) when $-[1+\frac{\sqrt{\frac{V}{U}}(V-U)}{8\pi G \ell^{2}\rho_{m}}]<w_{m}<-[\frac{1}{3}+\frac{\sqrt{\frac{V}{U}}(\frac{1}{3}V-U)}{8\pi G \ell^{2}\rho_{m}}-\frac{\Lambda}{12\pi G \rho_{m}}]$. Noteworthily, $S_{tot}^{(A)}$ is not only the matter's entropy but also the dark-energy's entropy provided by the cosmological constant $\Lambda$ and the spacetime self-coupling.

   \section{Conclusions and discussion}

   We generalized the Einstein Universe's the cosmical apparent horizon's properties and gravitational dynamics to the case of EBI Universe.
\par
   Firstly, we derived the holographic-style dynamical equations via the method in Ref.~\cite{14}, which combines the evolution of the Universe with the dynamics of the cosmical apparent horizon. To be specific, according to the value of $w_{m}$, we discussed the properties of the cosmical apparent horizon in EBI universe. Because the Universe is accelerated expanding at the present stage that means the matter outside the cosmical apparent horizon may enter into the cosmical apparent horizon with the evolution of the Universe. Hence we considered that the timelike apparent horizon is reasonable so that $-[1+\frac{1}{8\pi G\ell^{2}\rho_{m}}\sqrt{\frac{V}{U}}(V-U)]<w_{m}<[\frac{1}{3}+\frac{\Lambda}{6\pi G\rho_{m}}+\frac{1}{8\pi G\ell^{2}\rho_{m}}\sqrt{\frac{V}{U}}(\frac{1}{3}V+U)]$ is a rational range of the matter's EoS parameter, which give the correction compared to the previous range of $w_{m}$, $-1<w_{m}<\frac{1}{3}$ in Ref.~\cite{14}.
\par
    Via applying the Misner-Sharp energy and substituting the holographic-style dynamical equations, we obtained the total energy differential in the ($t, \Upsilon$) coordinates and ($t, r$) coordinates, respectively, corresponding to $dE=A[\rho_{m}+\frac{1}{8\pi G\ell^{2}}\sqrt{\frac{V}{U}}V+\frac{\Lambda}{8\pi G}]d\Upsilon-A[(\rho_{m}+p_{m})+\frac{1}{8\pi G\ell^{2}}\sqrt{\frac{V}{U}}(V-U)]\cdot H\Upsilon dt$ and $dE=A[\rho_{m}+\frac{1}{8\pi G\ell^{2}}\sqrt{\frac{V}{U}}V+\frac{\Lambda}{8\pi G}]adr-A[p_{m}-\frac{1}{8\pi G \ell^{2}}\sqrt{\frac{V}{U}}U-\frac{\Lambda}{8\pi G}]\cdot H\Upsilon dt$. We proved that these can be derived from the unified first laws of the gravitational dynamics $dE=A \boldsymbol{\tilde{\Psi}}+\tilde{W} d\hat{V}$ by redefining the total energy-momentum tensor $T_{\mu\nu}^{(tot)}$ instead of $T_{\mu\nu}^{(m)}$ in Hayward's approach~\cite{30} as well.
\par
    We also derived the total energy differential for the open system enveloped $\Upsilon_{A}$ :$dE_{A}=-\hat{T}_{A}dS_{A}+[\rho_{m}+\frac{\Lambda}{8\pi G}+\frac{1}{8\pi G\ell^{2}}\sqrt{\frac{V}{U}}V]d\hat{V}_{A}$, where the two extra terms $\frac{\Lambda}{8\pi G}d\hat{V}_{A}$ and $\frac{1}{8\pi G\ell^{2}}\sqrt{\frac{V}{U}}V d\hat{V}_{A}$ are corresponding to the dark energy provided by the cosmological constant and the spacetime self-coupling in EBI Universe, respectively. It illustrates that not only the matter's transition but also the dark-energy's fluxion arise on the cosmical apparent horizon $\Upsilon_{A}$ with the expansion of Universe.
\par
    Finally, we investigated the properties of the evolution of the total entropy $S_{tot}^{(A)}$ enclosed by the cosmical apparent horizon $\Upsilon_{A}$ in the EBI Universe. The results show that: when $-[1+\frac{\sqrt{\frac{V}{U}}(V-U)}{8\pi G \ell^{2}\rho_{m}}]<w_{m}<-[\frac{1}{3}+\frac{\sqrt{\frac{V}{U}}(\frac{1}{3}V-U)}{8\pi G \ell^{2}\rho_{m}}-\frac{\Lambda}{12\pi G \rho_{m}}]$ and the Universe is accelerated expanding ($\ddot{a}>0$), the generalized second laws of the nondecreasing entropy $S_{tot}^{(A)}$ is obtained. If $V$ satisfies the condition $V<3U-8\pi G\ell^{2}\rho_{m}\sqrt{\frac{U}{V}}$, $w_{m}$ can be a positive number to generate a accelerated expanding Universe. And we can obtain the generalized second laws of the nondecreasing entropy $S_{tot}^{(A)}$ without the dark matter when $V<3U-8\pi G\ell^{2}\rho_{m}\sqrt{\frac{U}{V}}$.
\par
    In a word, we proved that the EBI Universe has the similar properties of the apparent horizon dynamics and the gravitational thermodynamics like the Einstein Universe, and we also showed the differences between them.
    
\section*{Conflicts of Interest}
  The authors declare that there are no conflicts of interest regarding the publication of this paper.

\section*{Acknowledgments}
  We would like to thank the National Natural Science Foundation of China (Grant No.11571342) for supporting us on this work.

\section*{References}

\bibliographystyle{unsrt}
\bibliography{reference}

\end{document}